\documentclass{article}
\usepackage[sort&compress,square,comma,numbers]{natbib}

\usepackage{textcomp}

\usepackage{graphicx}  
\usepackage{dcolumn}   
\usepackage{bm}        
\usepackage{amssymb}   

\usepackage{soul,color}
\setstcolor{red}

\usepackage[strict]{changepage}

\newcommand{\uFOM}{{\textmu}FOM }

\begin{document}

\title{Brillouin cavity optomechanics with microfluidic devices}

\author{Gaurav Bahl$^{1\ast}$, Kyu Hyun Kim$^2$, Wonsuk Lee$^{2,3}$, \\Jing Liu$^3$, Xudong Fan$^3$, Tal Carmon$^2$\\
\footnotesize{$^1$Mechanical Science and Engineering, University of Illinois at Urbana-Champaign,}\\
\footnotesize{Urbana, Illinois, USA}\\
\footnotesize{$^2$Electrical Engineering and Computer Science, University of Michigan,}\\
\footnotesize{Ann Arbor, Michigan, USA}\\
\footnotesize{$^3$Biomedical Engineering, University of Michigan,}\\
\footnotesize{Ann Arbor, Michigan, USA}\\
\footnotesize{$^\ast$To whom correspondence should be addressed; E-mail: bahl@illinois.edu.}
}

\maketitle

\begin{abstract}
Cavity optomechanics allows the parametric coupling of phonon- and photon-modes in microresonators and is presently investigated in a broad variety of solid-state systems. Optomechanics with superfluids has been proposed as a path towards ultra-low optical- and mechanical-dissipation. However, there have been no optomechanics experiments reported with non-solid phases of matter. Direct liquid immersion of optomechanics experiments is challenging since the acoustic energy simply leaks out to the higher-impedance liquid surrounding the device. Conversely, here we confine liquids inside hollow resonators thereby enabling optical excitation of mechanical whispering-gallery modes at frequencies ranging from 2~MHz to 11,000~MHz (for example, with mechanical Q~=~4700 at 99~MHz). Vibrations are sustained even when we increase the fluid viscosity to be higher than that of blood.  Our device enables optomechanical investigation with liquids, while light is conventionally coupled from the outer dry side of the capillary, and liquids are provided by means of a standard microfluidic inlet.
\end{abstract}


Stimulated Brillouin scattering \cite{PhysRev.137.A1787,Yariv:1965ub} was considered for many years as an optical gain mechanism for lasers \cite{PhysRevLett.12.592} and for nonlinear optics including phase conjugation \cite{Zeldovich72} and slow light \cite{PhysRevLett.94.153902}. Subsequently, the platforms for Brillouin scattering were extended from bulk materials and fibers, to droplets \cite{Zhang:89}, nano-spheres \cite{PhysRevLett.90.255502}, photonic-crystal fibers \cite{Dainese:2006p1365} and crystalline resonators \cite{GrudininCaF2lasing,Savchenkov:11}. The recent demonstration of Brillouin scattering in microspheres \cite{Tomes2009} was followed by the Brillouin cooling \cite{BahlNP2012} and excitation \cite{Savchenkov:11,Bahl:2011cf} of their vibrational modes as well; which indicates that Brillouin effects can serve as a general mechanism for actuating (and interrogating) vibration in various types of micro-mechanical resonators. In separate research on optofluidic devices \cite{Psaltis:2006fe,stolyarovNatPhoton2012}, the motion of liquids has been used to control light, but light has rarely been used to actuate a fluid. In this study, we use the Brillouin scattering of light from sound to excite and measure vibration in a microfluidic optomechanical ({\textmu}FOM) resonator. We design, fabricate, and actuate \uFOM resonators that exhibit mechanical deformation at their solid-fluid interface. Here, confinement of the test fluid inside the device mitigates acoustic loss. Conversely, if an optomechanical resonator will just naively be submerged in a liquid, acoustic radiative losses will increase on account of the acoustic impedance of liquids being much larger than that of air.

\begin{figure}[h]
	\centering
		\includegraphics[width=0.8\hsize,clip=true, trim=3.1in 3.4in 3.7in 0.4in]{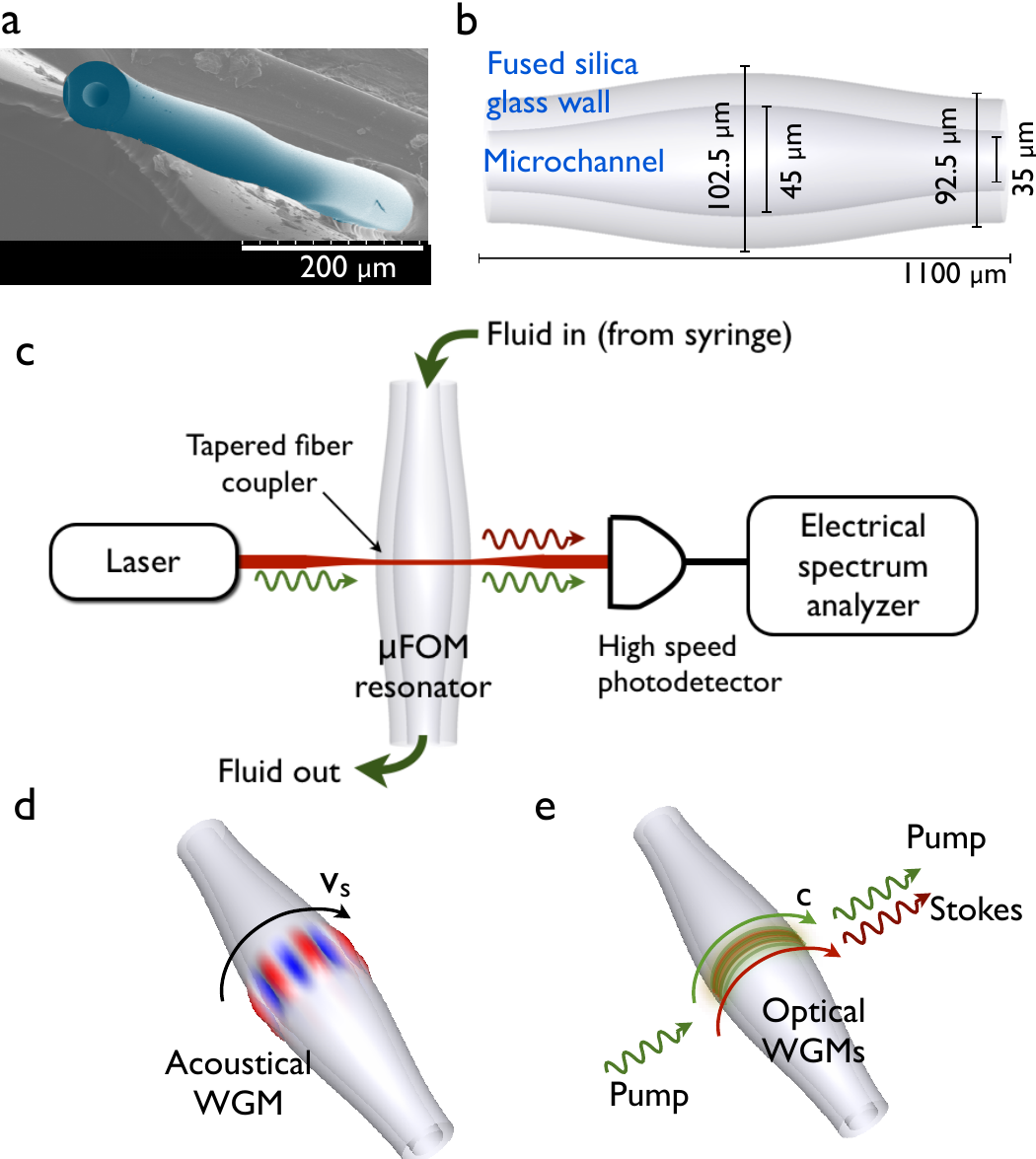}
	\caption{
	\textbf{(a)} Colorized scanning electron micrograph of a fused silica \uFOM device, showing the modulation of the radius as a function of position. 
	\textbf{(b)} Dimensions of the device used to obtain mechanical whispering-gallery modes in Fig. \ref{fig:ExperimentalModes}b-f. Not drawn to scale.
	\textbf{(c)} A tapered optical fiber is used to couple light in and out of the whispering-gallery optical modes of the \uFOM resonator (without contact). We employ a telecom wavelength $1.5$ {\textmu}m pump laser without any modulation. A high speed photodetector measures the mechanical vibration by means of its optical signature, i.e. the beat note between the optical pump and the Stokes scattered light. The experimental configuration for forward scattering is illustrated here.
	\textbf{(d)} Illustration of the mechanical whispering-gallery modes on the \uFOM resonator, showing that mechanical displacement is concentrated at the equator. The mode circulates at the velocity of sound, $v_s$.
	\textbf{(e)} Illustration of the optical whispering-gallery modes of the \uFOM resonator that travel at the speed of light, $c$.
	\label{fig:ExperimentalSetup}}
\end{figure}

\begin{figure*}[htb]
	\centering
		\includegraphics[width=\hsize, clip=true, trim=0.5in 5.8in 1.2in 0.7in]{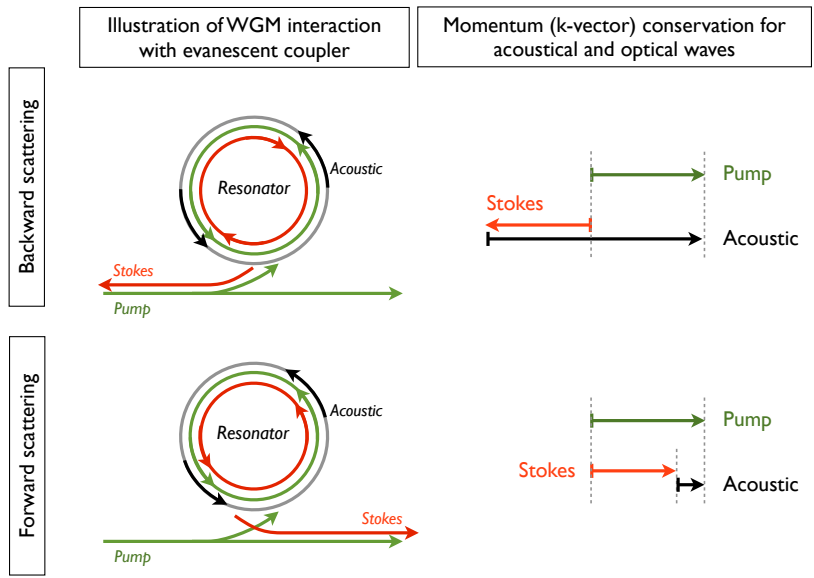}
	\caption{
	\textbf{Momentum conservation} dictates the acoustic WGM frequencies that are generated through stimulated Brillouin scattering. Light traveling in the tapered waveguide evanescently couples to optical WGMs of the resonator, and is scattered to Stokes frequencies in either the forward or backward direction.
	\textbf{In the case of back-scattering (B-SBS)} momentum conservation enforces long acoustic k-vectors such that the acoustic modes are at high frequencies (i.e. 10 GHz regime). Stokes light is received from the same tapered coupler in the backward direction.
	\textbf{In forward scattering (F-SBS)} momentum conservation enforces that the acoustic modes are at much lower frequencies (i.e. $<$ 1 GHz regime). Stokes light is received in the forward direction as illustrated in Fig.~\ref{fig:ExperimentalSetup}c.
	\label{fig:kVector}}
\end{figure*}

\begin{figure}[h]
	\begin{adjustwidth}{-1in}{-1in}%
	\centering
		\includegraphics[width=1.4\hsize,clip=true, trim=2in 4.1in 2in 0.8in]{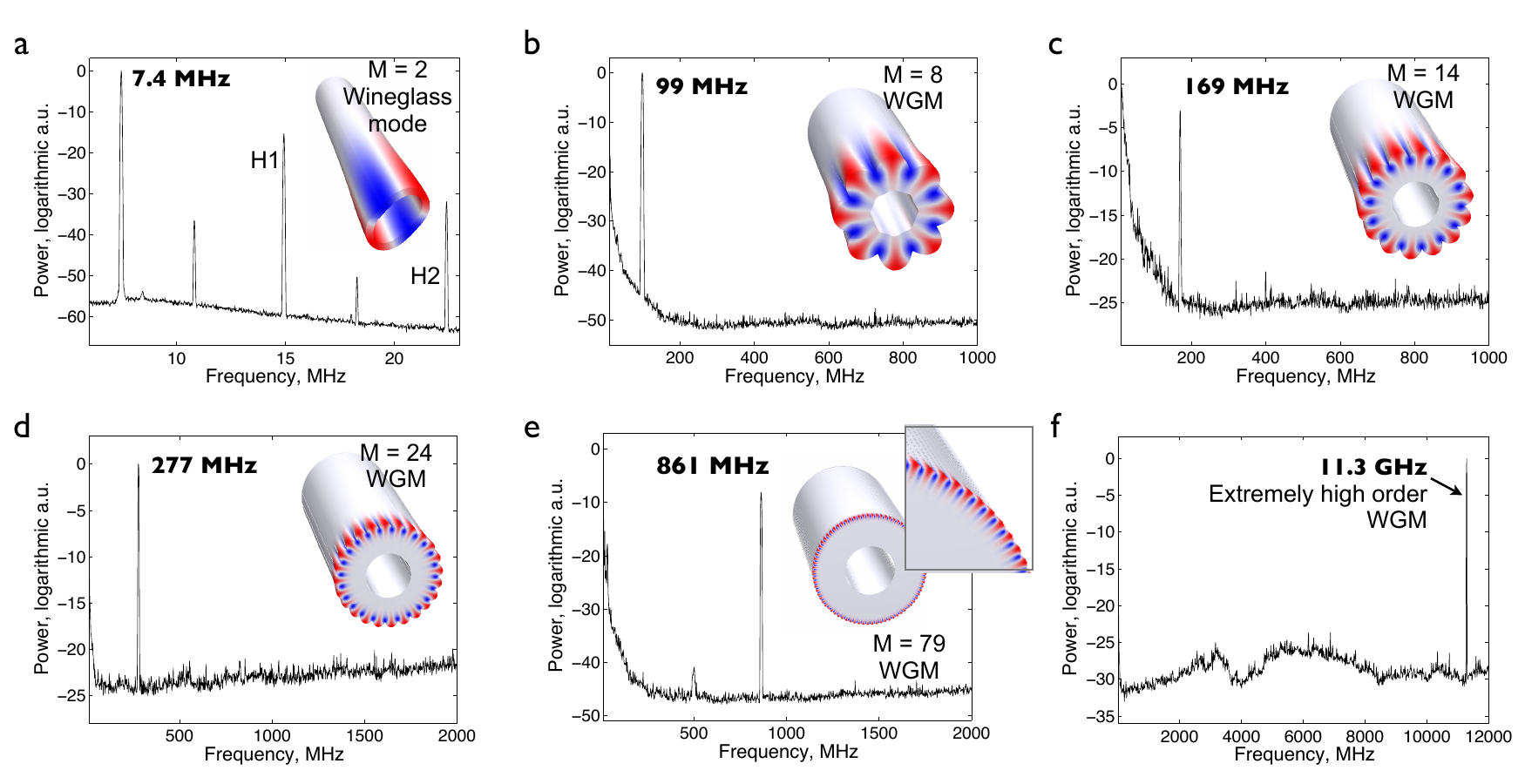}
	\caption{\textbf{Experimental observation of mechanical modes} in water-filled devices. 
	\textbf{(a)} Wineglass mechanical mode at 7.4 MHz. H1 and H2 are the first and second harmonics respectively. Frequency spurs at 10.81 MHz and 18.27 MHz are caused by other mechanical modes of vibration.
	\textbf{(b), (c), (d), (e)} Acoustic whispering-gallery modes of various azimuthal mode orders identified with the number $M$. These modes are excited by forward stimulated Brillouin scattering.
	\textbf{(f)} Extremely high order acoustic WGMs near 11 GHz are excited by backward stimulated Brillouin scattering.
	\label{fig:ExperimentalModes}}
	\end{adjustwidth}
\end{figure}

Our \uFOM device (Fig. \ref{fig:ExperimentalSetup}a, b) is fabricated from a lengthwise-drawn fused silica capillary \cite{Sun:2010cn} with its radius modulated as a function of length (see Methods). 
At its widest point this glass device forms a whispering-gallery mode (WGM) microresonator \cite{ChangMicrocavities} that resonantly enhances both light and sound (Fig.~\ref{fig:ExperimentalSetup}d,e) while supporting a considerable overlap between the optical and mechanical modes. The acoustic modes are optically excited by means of forward \cite{Bahl:2011cf,Savchenkov:11,BahlNP2012} and backward \cite{GrudininCaF2lasing,Tomes2009} stimulated Brillouin scattering, i.e. F-SBS and B-SBS respectively.

In the case of B-SBS \cite{Boyd}, the incoming pump light (frequency $\omega_p$) photoelastically back-scatters from a high-frequency acoustic wave (frequency $\Omega$) in the material to generate a lower frequency Stokes optical signal (frequency $\omega_s = \omega_p - \Omega$) as shown in Fig.~\ref{fig:kVector}. Simultaneously, electrostrictive pressure is generated by the two optical signals which amplifies the acoustic wave. Optical spectrum analysis reveals the optical pump and Stokes signals, as shown previously in \cite{Tomes2009}. The resulting beat note between pump and Stokes signals measured on an electrical spectrum analyzer is representative of the high-frequency acoustic wave. This B-SBS positive feedback process occurs near $\Omega = 11$~GHz mechanical frequency in silica when 1.5 {\textmu}m pump light is used \cite{Boyd}. An experimental example of such a mode in this study is shown in Fig. \ref{fig:ExperimentalModes}f.

The case for forward scattering, i.e. F-SBS, is identical to what is described above, except that the pump light photoelastically scatters in the forward direction such that the Stokes signal co-propagates with the pump. This reversal of the scattering direction (to forwards) suggests that much smaller phonon momentum is required to bridge between the pump and Stokes momentum gap (Fig.~\ref{fig:kVector}). As a result, lower acoustic frequencies ($<$ GHz) are more likely for the F-SBS process.
Additional details on the energy and momentum conservation requirements \cite{Yariv} for F-SBS and B-SBS in a circular resonator are discussed in \cite{GrudininCaF2lasing, MatskoSAWPRL, Zehnpfennig:11, Savchenkov:11, Tomes2009,BahlNP2012,Bahl:2011cf,Shelby:1985p1169}. 
Specifically for the capillary geometry that we use here, it is indicated \cite{BahlNJP12} that a wide variety of acoustic whispering-gallery modes exist in the shell-type geometry.

\section*{Results}

\textbf{Experimental excitation of acoustic WGMs in fluid-filled \uFOM resonators: }\\
We experimentally excite acoustic whispering-gallery modes (WGM) in a water-filled \uFOM resonator, ranging in frequency from 99 MHz to 11,000 MHz (Fig. \ref{fig:ExperimentalModes}b-f). As described in Fig. \ref{fig:kVector}, the ``Brillouin'' optomechanical process \cite{Bahl:2011cf} allows for excitation and measurement of these acoustic WGMs by coupling continuous-in-time light at an optical resonance, while interrogating the optomechanical oscillation at the fiber output of the system (see Methods). No modulation of the pump laser is needed.
We introduce and extract light from the optical WGMs by means of evanescent coupling with a tapered optical fiber \cite{Knight:1997p1319,Cai:01} (as shown in Fig. \ref{fig:ExperimentalSetup}c). Optical ring-down measurements \cite{Savchenkov:07} indicate optical quality factors in excess of $Q_o=10^8$ (highest measured $Q_o = 1.6 \times 10^8$). No contact between the fiber and the \uFOM resonator is required for mechanical transduction, so the acoustic quality factors are not degraded. The same optical fiber taper is used to extract the Stokes scattered light from the device, which is then used as a measure of the mechanical vibration within the structure \cite{BahlNP2012}. 
All of our experiments are performed at room temperature and atmospheric pressure, and various solutions are introduced into the device by means of a syringe pump.

We calculate the mechanical modes of the \uFOM resonator using 3D finite element modeling \cite{Zehnpfennig:11,Comsol, BahlNJP12}. Each of these modes is identified with a M-number which is the azimuthal mode order quantifying the number of acoustic wavelengths around the equatorial circumference of the device. In addition, a M=2 whispering-gallery mode at 7 MHz is observed (Fig. \ref{fig:ExperimentalModes}a), that is generally referred to as a wineglass mode.
The harmonics in this wineglass mode (H1, H2 in Fig.\ref{fig:ExperimentalModes}a) are typical for such low frequency modes and result from the fact that this is a standing-wave type vibration \cite{Carmon2005,Rokhsari:05,PhysRevLett.95.033901,Kyu_NP}, as opposed to the traveling wave vibrations in Fig.\ref{fig:ExperimentalModes}b-f (more details in \cite{BreatheWineglassNote1}).
\\

\begin{figure}[htb]
	\centering
		\includegraphics[width=0.50\hsize,clip=true, trim=2.5in 2.8in 6in 1in]{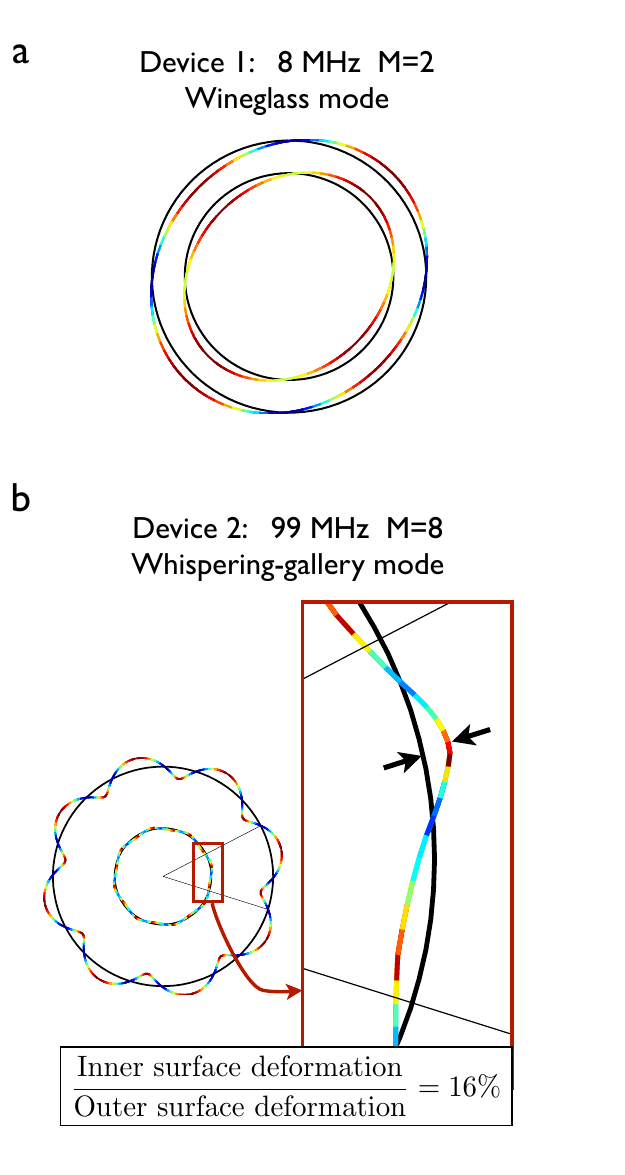}
	\caption{\textbf{Calculated equatorial mechanical mode profiles.} 
	\textbf{(a)} For the M=2 wineglass mode at 8 MHz (experimentally appears at 7.4 MHz, see Fig. \ref{fig:ExperimentalModes}a) the displacement ratio of internal and external surfaces is almost 1:1 at the equator.
	\textbf{(b)} For the experimentally observed M=8 acoustic WGM at 99 MHz (see Fig. \ref{fig:ExperimentalModes}b), the inner surface experiences 16\% displacement relative to the outer surface.  
	\label{fig:InnerDisp}}
\end{figure}

\textbf{Quantifying solid-liquid interaction: }\\
The penetration of sound to the liquid scales with the ratio between the acoustic wavelength and the thickness of the resonator wall. Therefore the 7.4~MHz (M=2) mode is calculated to have large penetration into the fluid (Fig.~\ref{fig:InnerDisp}a). Finite element calculation shows that the 99 MHz (M=8) mode has a 16\% penetration to fluid, defined by the deformation amplitude at the inner fluid interface divided by the maximum deformation (Fig. \ref{fig:InnerDisp}b). The higher M modes have lower penetration into the liquid. However, as we can fabricate the resonator with walls as thin as 560 nm \cite{Wonsuk2011} we believe that a large acoustic penetration to water will be possible in such thin-wall silica bubble resonators even for the 11~GHz mode (M $\approx$ 650), 
which can open a rare hypersonic window for the acoustic analysis of liquids. Reduction in wall thickness may even allow access to an interesting regime where giant-enhanced Brillouin scattering is predicted \cite{PhysRevX.2.011008}.\\

\begin{figure}[htb]
	\centering
		\includegraphics[width=0.8\hsize,clip=true, trim=2.6in 1.5in 4.33in 0.3in]{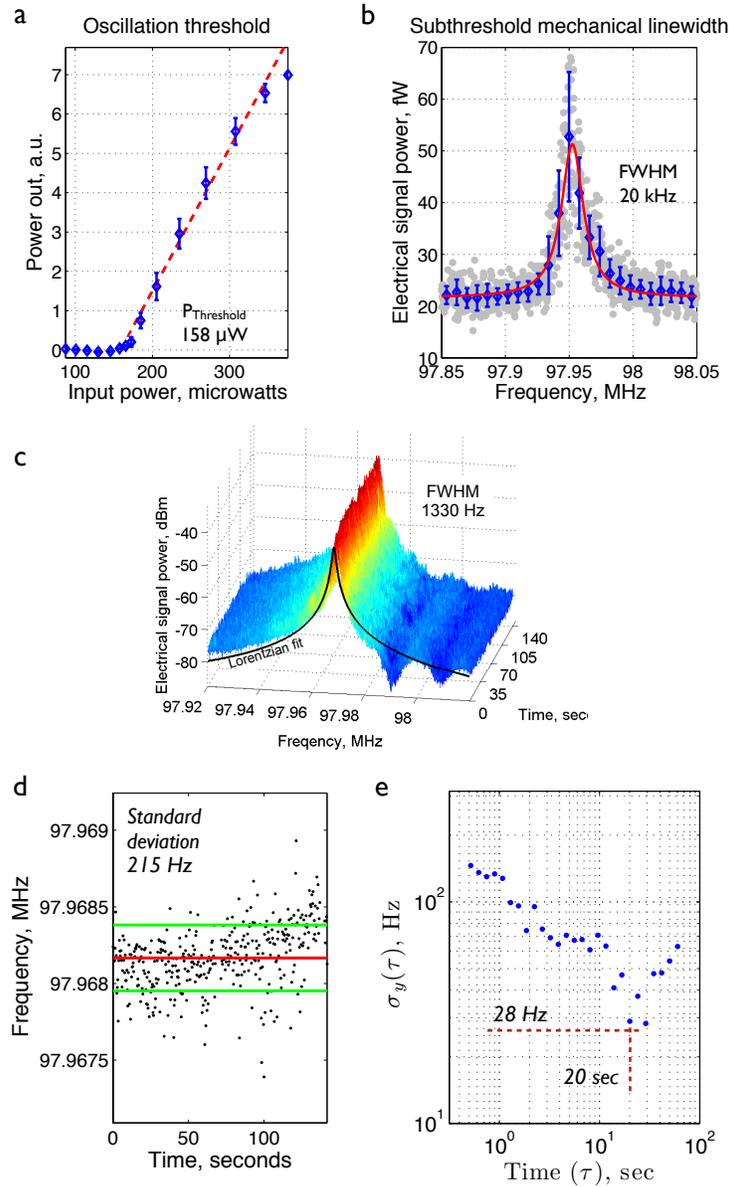}
	\caption{\textbf{Mode and oscillator characterization} for the 99 MHz acoustic WGM in a water-filled device. 
	\textbf{(a)} Lowest observed optical threshold power to excite mechanical oscillation is $158$ {\textmu}W. 
	\textbf{(b)} The intrinsic mechanical linewidth (mechanical quality factor) is directly observed at very low input optical power.
	\textbf{(c)} A spectrogram obtained over 140 seconds while scanning the laser over the optical resonance that excites this mechanical vibration. 
	\textbf{(d)} Peak frequencies extracted from spectrogram of (c) show a standard deviation of 215 Hz.
	\textbf{(e)} Allan deviation plot of the frequency data in (d) provides an estimate of the potential sensor resolution.
	\label{fig:LongExperiment}}
\end{figure}

\textbf{Characterization of mechanical mode: }\\
We measure that the minimum power required to excite these oscillations in the water-filled \uFOM device is 158 {\textmu}W (Fig. \ref{fig:LongExperiment}a). We also measured a mechanical quality factor $Q_m = 4700$ for the 99 MHz (M=8) mode in this water-filled device (Fig. \ref{fig:LongExperiment}b), via the sub-threshold oscillation linewidth, which implies $Q_m \times \textrm{Frequency} = 4.6 \times 10^{11}$. This ``f-Q product'' compares well against the previously measured $Q_m=12\,300$ in a solid silica sphere for a 95~MHz mode \cite{BahlNP2012}. 

This 99 MHz (M = 8) mode of vibration can be maintained as long as the CW input laser power is provided. A spectrogram demonstrating this stability (Fig. \ref{fig:LongExperiment}c) is obtained over 140 seconds while periodically scanning through the optical resonance that excites this mechanical mode. We detect the peak oscillation frequencies (Fig. \ref{fig:LongExperiment}d) in this spectrogram by means of Lorentzian curve fits (Fig. \ref{fig:LongExperiment}c) at each time-point. The standard deviation of the peak frequency is calculated to be 215 Hz along this 140-second period. In order to better characterize the stability of the oscillation and the ability of this system to function as a sensor we calculate the Allan deviation (square root of the two-point variance) of the data, which charts the frequency deviation as a function of averaging time (Fig. \ref{fig:LongExperiment}e). Here we see that a frequency resolution of 28 Hz is achievable by utilizing a moving average over 20 seconds. \\

\begin{figure}[htb]
	\centering
		\includegraphics[width=0.7\hsize,clip=true, trim=0.5in 7.3in 4.8in 0in]{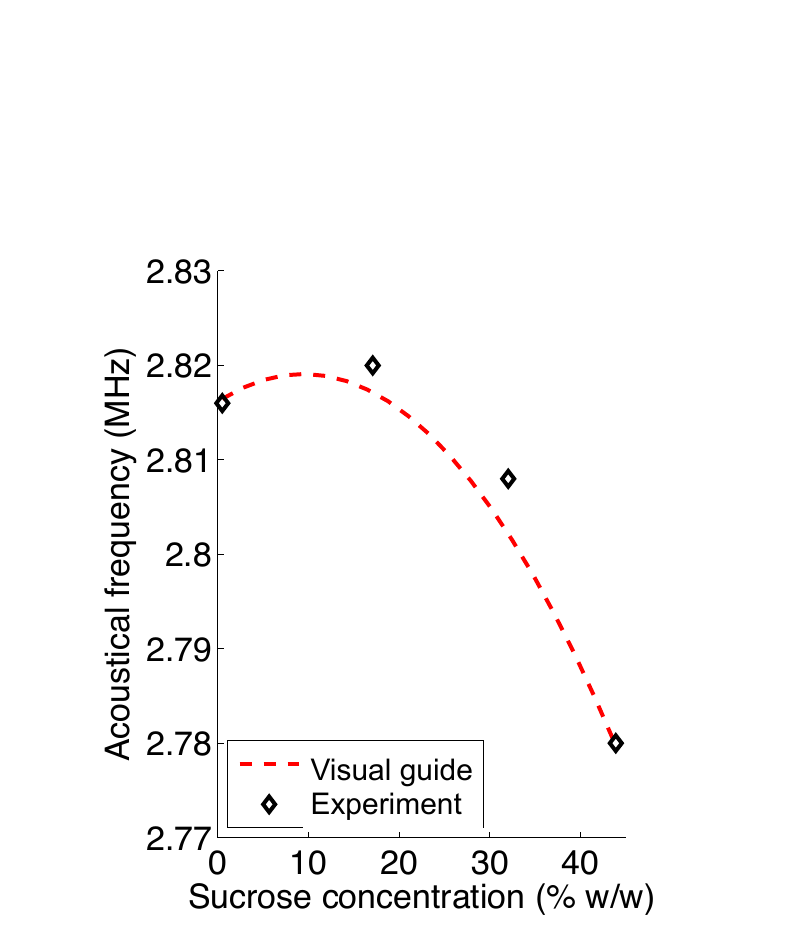}
	\caption{\textbf{Optomechanical sensing of sucrose solutions.} 
	We experimentally measure the frequency of a 2.8~MHz wineglass acoustic mode (M=2) as the concentration of sucrose solution inside the \uFOM resonator is varied. The dashed red line is a quadratic guide for the eye. Resonator outer diameter is 184 {\textmu}m.
	\label{fig:FrequencyTheory}}
\end{figure}

\textbf{Operation with high-density high-viscosity liquids:}\\
We perform an experiment to prove that the opto-mechanical interaction can be sustained even when the motional mass is high (i.e. high density liquid) and when the fluid-related acoustic energy losses are high (i.e. high viscosity).
We incorporate sucrose (table sugar) into an aqueous solution inside the resonator and measure the sensitivity of the optomechanically actuated wineglass mode to the sucrose concentration of the fluid. 
Optomechanical oscillation is observed at all tested sucrose concentrations.
The resulting relationship between acoustic frequency and the sucrose concentration exhibits a non-monotonic trend, which we plot in Fig.~\ref{fig:FrequencyTheory} along with a trend line to guide the eye.
This non-monotonic trend 
might originate from non-perturbative high sugar concentrations resulting in nonlinear change in frequency.
We note that at the highest sucrose concentration that we test in this experiment, the solution has a viscosity that is approximately 3-times greater than that of blood \cite{Iscotables}.

\section*{Discussion}

In this work we demonstrate a bridge between optomechanics and microfluidics by confining various liquids inside a hollow microfluidic optomechanical resonator. We present experimental evidence that the optomechanical interaction in the \uFOM device is dependent on the fluid contained within. These results are therefore a step towards novel experiments probing optomechanics on non-solid phases of matter.
In particular, the high frequency, high quality-factor mechanical modes demonstrated in this work may enable strongly localized, high-sensitivity, optomechanical interaction with chemical and biological analytes \cite{Burg2007,BartonCraighead} since environmental parameters like pressure and cell nutrients are relatively easy to control inside the hollow resonator \cite{Sumetsky:10,Wonsuk2011,Berneschi:11}, and since the liquid volume within the device is the scale of a living cell.

In contrast to other optomechanical systems that typically involve standing-wave mechanical vibrations, here circulating phonons comprise the acoustic WGMs \cite{Bahl:2011cf,Zehnpfennig:11}, which are essentially acoustic vortices \cite{Vortex1983,LalVortex,PhysRevLett.96.043604}. These traveling-wave type acoustic modes carry angular momentum, which creates possibility of using \uFOM devices to enable optomechanical interactions with vortices in various liquids \cite{Feynman195517,Roberts:1974wd,PhysRevLett.39.1208} and condensates.
\\

\section*{Methods}

\subsection*{Microfluidic device fabrication}
The \uFOM device (Fig. \ref{fig:ExperimentalSetup}a,b) is fabricated with a fused-silica glass capillary that is pulled lengthwise using linear actuators, while being heated with an infrared laser (CO$_2$ at 10.6 {\textmu}m) as previously described in \cite{Sun:2010cn}. By controlling the power of the manufacturing laser in the process of pulling, we can control the device diameter along its length. Optical WGMs as well as acoustic WGMs are simultaneously well-confined in the regions of large diameter sandwiched between regions of narrow diameter (Fig.~\ref{fig:ExperimentalSetup}d,e), enabling a high-degree of modal overlap. Therefore, these large diameter regions form the \uFOM resonators, and multiple such resonators can be built on a single capillary.

\subsection*{Experimental method}
The \uFOM resonator is placed in close proximity ($<$ 1 \textmu{m} distance) to tapered optical fiber \cite{Knight:1997p1319,Cai:01} (as shown in Fig. \ref{fig:ExperimentalSetup}c), such that evanescent coupling from the taper allows light to be introduced and extracted from the optical WGMs of the device. A CW fiber-coupled tunable 1.5 \textmu{m} laser is used as the source (Newfocus TLB-6328), and a fiber-coupled high-speed photodetector is used to monitor vibration. Mechanical vibration is stimulated within the device by pumping specific optical resonances where the three-mode phase-match exists. The internally generated Stokes-shifted laser is then also collected by the tapered fiber, and the temporal interference of this new Brillouin laser against the partially transmitted pump laser is used to infer the mechanical response using the photodetector \cite{Bahl:2011cf,BahlNP2012}. An optical spectrum analyzer is used to verify that 4-wave mixing and Raman scattering are not responsible for the beat note observed in the electrical spectrum.

\section*{Acknowledgements}
This work was supported by the Defense Advanced Research Projects Agency (DARPA) Optical Radiation Cooling and Heating in Integrated Devices (ORCHID) program and through a grant from the Air Force Office of Scientific Research (AFOSR).



\begin{thebibliography}{10}
\bibitem{BreatheWineglassNote1}
{Uniquely for this 7 MHz mode, we measure high harmonics of the fundamental
  frequency. As has been previously calculated and measured [refs. 28-31], such
  harmonics indicate that the acoustic mode is of standing-wave type, and is
  excited by centrifugal radiation pressure.}
\end{thebibliography}
\end{document}